\documentclass[aps,twocolumn,superscriptaddress]{revtex4-2}

\usepackage[english]{babel}
\usepackage{graphicx}
\usepackage{amsmath, amssymb, amsfonts, amsthm, mathrsfs, bm}
\usepackage{color}
\usepackage{hyperref}
\hypersetup{colorlinks=true, citecolor=blue, urlcolor=blue, linkcolor=blue}
\usepackage{cleveref}
\usepackage{lipsum}

\begin{document}

\title{Particle deformability stabilizes hexatic order and suppresses crystallization}
\author{Jatin Kumar}
\affiliation{School of Physical and Mathematical Science, Nanyang Technological University, Singapore}
\author{Wu Zeng}
\affiliation{School of Physical and Mathematical Science, Nanyang Technological University, Singapore}
\author{Anshuman Pasupalak}
\affiliation{School of Physical and Mathematical Science, Nanyang Technological University, Singapore}
\author{Massimo Pica Ciamarra}
\affiliation{School of Physical and Mathematical Science, Nanyang Technological University, Singapore}
\affiliation{
CNR--SPIN, Dipartimento di Scienze Fisiche,
Universit\`a di Napoli Federico II, I-80126, Napoli, Italy
} 
\date{\today}
\newcommand{\MPC}[1]{{\color{red}{\textbf{#1}}}}
\newcommand{\mpc}[1]{{\color{red}{\textbf{#1}}}}
\newcommand{\JK}[1]{{\color{blue}{\textbf{#1}}}}

\begin{abstract}
We show that two-dimensional systems of deformable particles undergo a continuous liquid–hexatic transition upon compression or cooling, but no hexatic–solid transition—even at zero temperature and high density. 
Numerical simulations reveal that solid-like configurations do not possess a lower energy than hexatic ones, so that at low temperatures the hexatic phase is thermodynamically favored due to its higher entropy. 
Dislocation condensation, necessary for solid formation, is suppressed as the system accommodates strain via particle shape changes, responding affinely to compression.
Our findings identify a generic route by which microscopic mechanical properties control defect energetics and reshape phase behavior in two dimensions, with broad relevance for soft and biological materials such as microgels and epithelial tissues.
\end{abstract}

\maketitle

Melting in two dimensions remains a central problem in statistical physics. 
For systems with short-range interactions, the Mermin--Wagner theorem forbids true long-range positional order, yet solids with quasi-long-range translational order and long-range bond-orientational order can still exist. 
The Kosterlitz--Thouless--Halperin--Nelson--Young (KTHNY) theory~\cite{kosterlitz1973ordering,halperin1978theory,young1979melting} predicts that such solids melt via two continuous defect-mediated transitions: unbinding of dislocations produces a hexatic phase with algebraic orientational correlations, and subsequent proliferation of disclinations yields a liquid.

Experiments~\cite{Rosenbaum1983,Zahn1999,Maret2004,Dillmann2012,Thorneywork2017} 
and simulations~\cite{bernard2011two,kapfer2015two,zu2016density,Anderson2017,li2018role,Li2020} 
have broadly supported this framework while also revealing important variations, including first-order liquid--hexatic transitions and cases where the hexatic phase is suppressed. 
A robust empirical feature, however, is the close link between hexatic and solid order: whenever a hexatic phase appears, a solid phase is recovered at higher density or lower temperature, unless quenched disorder or polydispersity suppress both phases simultaneously~\cite{sampedro2019melting,John_Russo,Li2023}.

Many soft and biological materials---such as microgels, emulsions, and epithelial tissues---consist of particles that deform appreciably under stress. 
Deformability plays a central role in mediating mechanical response and collective organization, as suggested by models of deformable particles and by studies of compressed microgels.
Whether such deformability preserves, modifies, or breaks the conventional connection between hexatic and solid phases remains unknown.

Here we show that particle deformability fundamentally reshapes two-dimensional phase behavior. 
Using molecular dynamics simulations of a deformable-particle model~\cite{boromand2018jamming}, we find a stable hexatic phase that persists down to zero temperature and up to high density, without ever crystallizing.
Cooling or compression never induces translational order. 
Instead, deformability produces an affine mechanical response that relaxes local stresses within each particle, suppressing the stress localization required for dislocation binding and thereby eliminating the solid phase entirely. 
As a result, the hexatic phase becomes both entropically favorable and energetically competitive even in the $T\to0$ limit, in contrast with the conventional entropy-only stabilization mechanism in hard-particle systems and consistent with the recovery of ordinary melting behavior in stiffer deformable-particle models~\cite{Guo2023}.

These results reveal a generic mechanism by which microscopic mechanical properties---specifically particle deformability---control defect energetics and phase stability in two dimensions, suggesting that soft and biological materials may generically realize phase behavior that departs from the classical KTHNY framework.

\begin{figure}[!t]
  \centering
  \includegraphics[width=0.48\textwidth]{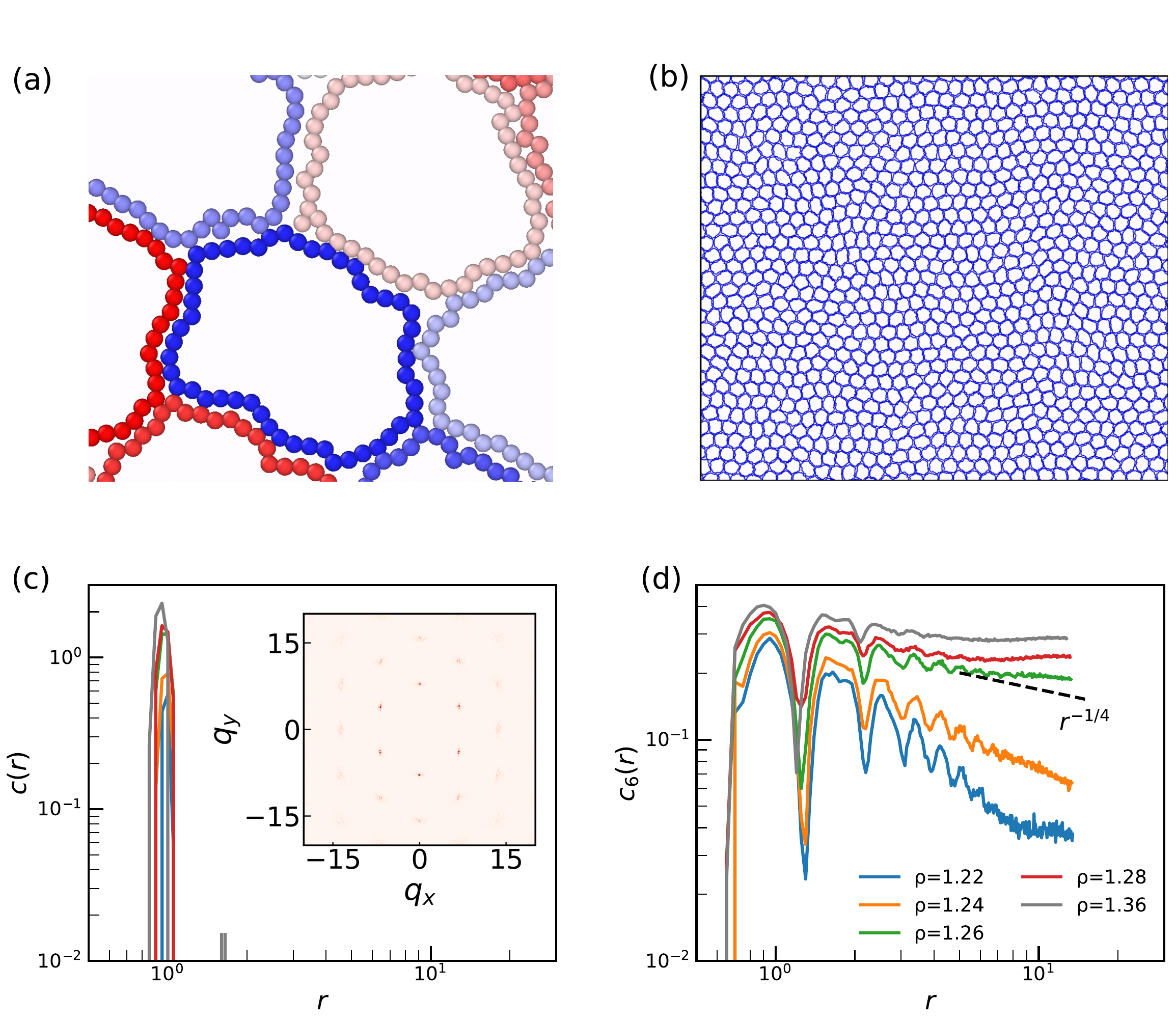}    
  \caption{
  (a) Zoomed-in view of the investigated system, highlighting the polymer-ring description of the particles, represented in different colors, and (b) configuration of the system at $\rho = 1.32$ and $T = 0.1$.
  Translational (c) and bond-orientational (d) correlation function at $T = 0.1$, for diverse values of the density. The translational correlation function $c(r)$ decays exponentially at all densities, indicating that the system does not enter the solid phase. In contrast, $g_6(r)$, which decays exponentially at low densities, exhibits a slow power-law decay for $\rho \gtrsim 1.26$, signaling the transition from the liquid to the hexatic phase. 
  The inset of (c) illustrates the static structure factor at $\rho = 1.36$. The smearing of the first peaks is consistent with the loss of translational order.
  The data in (c) and (d) are averaged over 10 independent realizations.
  \label{fig:Tcorrelations}
  }
\end{figure}

To explore how deformability alters defect energetics and phase stability, we investigate the phase behavior of the deformable particle model introduced in Ref.~\cite{boromand2018jamming}. 
Each particle is described as a ring polymer composed of $n$ monomers, subject to a constraint on its enclosed area, as illustrated in Fig.~\ref{fig:Tcorrelations}. 
This model uniquely captures both low-density regimes, where particles behave as isolated cells, and high-density regimes, where they become confluent.
It thus differs fundamentally from other models of deformable particles that enforce confluence—such as the Voronoi and vertex models—which have been shown to exhibit hexatic and solid phases~\cite{li2018role,pasupalak2020hexatic}.
The energy of an isolated particle is given by $E = \sum_{i=1}^{n} k_l (l_i - l_0)^2 + k_a (A - A_0)^2$, where $l_i$ are edge lengths, $A$ is the cell area, and $k_l$, $k_a$ are elastic constants. The parameters $l_0$ and $A_0$ denote the preferred edge length and area, respectively. Monomers from different particles interact via the purely repulsive Weeks--Chandler--Andersen potential, defined for $r \leq d_m = 2^{1/6}\sigma$ as $u(r) = 4\epsilon \left[ \left( \frac{\sigma}{r} \right)^{12} - \left( \frac{\sigma}{r} \right)^6 \right]$, where $r$ is the inter-monomer distance, $\sigma$ the monomer diameter, and $\epsilon$ the interaction strength.
We fix the number of monomers to $n = 50$, set $n\epsilon = 1$, $k_l = 10^6\,\epsilon$, $l_0 = \pi/n$, $k_a = 5 \cdot 10^4\,\epsilon$, and $2^{1/6}\sigma = 1.1\,l_0$. 
In this model, the mechanical properties of isolated cells are controlled by the dimensionless shape parameter $p_0 = nl_0/\sqrt{A_0}$, which we set to $p_0 = 3.9$ to ensure high deformability. 
This choice determines the preferred area $A_0$ and the typical length scale $d_0 = 2\sqrt{A_0/\pi}$, which we use as our unit of length. 
To probe equilibrium properties, we perform molecular dynamics simulations of 
$N= 1024$ cells, with thermal fluctuations introduced via a Langevin thermostat applied to each cell’s center of mass.
We verify the absence of finite-size effects in the End Matter.

To distinguish the various pure phases, we examine the spatial decay of both the translational correlation function, $c(r)$, and the bond-orientational correlation function, $c_6(r)$, where $r = |\vec{r}_i - \vec{r}_j|$ is the distance between the centers of mass of particles $i$ and $j$. 
The translational correlation is defined as $c(r) = \langle e^{i \vec{G} \cdot (\vec{r}_i - \vec{r}_j)} \rangle$, where $\vec{G}$ denotes one of the primary reciprocal lattice vectors obtained from the static structure factor $S(\vec{q}) = \frac{1}{N} \left| \sum_{j=1}^{N} e^{i \vec{q} \cdot \vec{r}_j} \right|^2$ at the considered state point~\cite{bernard2011two,li2019accurate}. 
The bond-orientational correlation function is given by $c_6(r) = \langle \psi_6(\vec{r}_i)\, \psi_6^*(\vec{r}_j) \rangle$, where $\psi_6(\vec{r}_j) = \frac{1}{n_j} \sum_{m=1}^{n_j} \exp(i\,6\,\theta_j^m)$ is the local sixfold bond-orientational order parameter. 
The sum runs over the $n_j$ Voronoi neighbors of particle $j$, and $\theta_j^m$ is the angle between the bond vector $(\vec{r}_m - \vec{r}_j)$ and a fixed reference axis.

As shown in Fig.~\ref{fig:Tcorrelations}(c) and (d), both correlation functions decay rapidly at low density, consistent with a liquid phase. 
For $\rho \gtrsim 1.26$, $c(r)$ continues to decay exponentially, while $c_6(r)$ exhibits a power-law decay, $c_6(r) \propto r^{-\eta_6}$ with $\eta_6 < 1/4$, identifying the system as hexatic~\cite{kosterlitz1973ordering,halperin1978theory,young1979melting}. 
Surprisingly, further compressing the system to the maximum density at which it becomes unstable due to particle buckling does not affect $c(r)$; compression does not induce translational order, so the system persists in the hexatic phase. 
This absence of translational order is further supported by the structure factor, shown in the inset of Fig.~\ref{fig:Tcorrelations}(c) at the highest simulated density: the first diffraction peaks remain broad and diffuse, rather than sharpening into Bragg-like peaks, and the second-shell peaks are strongly suppressed. 
Such a structure factor is incompatible with quasi-long-range positional order and is fully consistent with a persistent hexatic phase.

\begin{figure}[t!]
  \centering
  \includegraphics[width=0.48\textwidth]{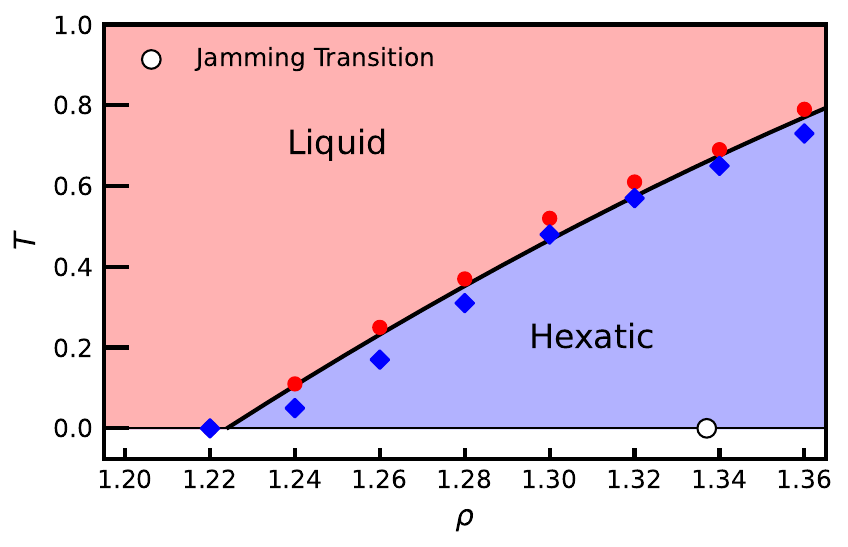}
  \caption{Equilibrium phase diagram for systems of deformable particles. 
  At each density, diamonds mark the highest temperature at which we have estimated the system to be in the hexatic phase, and circles mark the lowest temperature of the liquid phases. The line is a guide to the eye. The white circle denotes the jamming transition density estimated by compressing the system at $T=0$.
  }
  \label{phase_diagram}
\end{figure}

By repeating this analysis across a range of temperatures, we construct the equilibrium phase diagram shown in Fig.~\ref{phase_diagram}. 
The diagram reveals a temperature-dependent liquid-to-hexatic transition, but no subsequent hexatic-to-solid transition—neither upon cooling nor compression—up to the highest investigated density.

\begin{figure}[!t]
\centering
\includegraphics[width=0.48\textwidth]{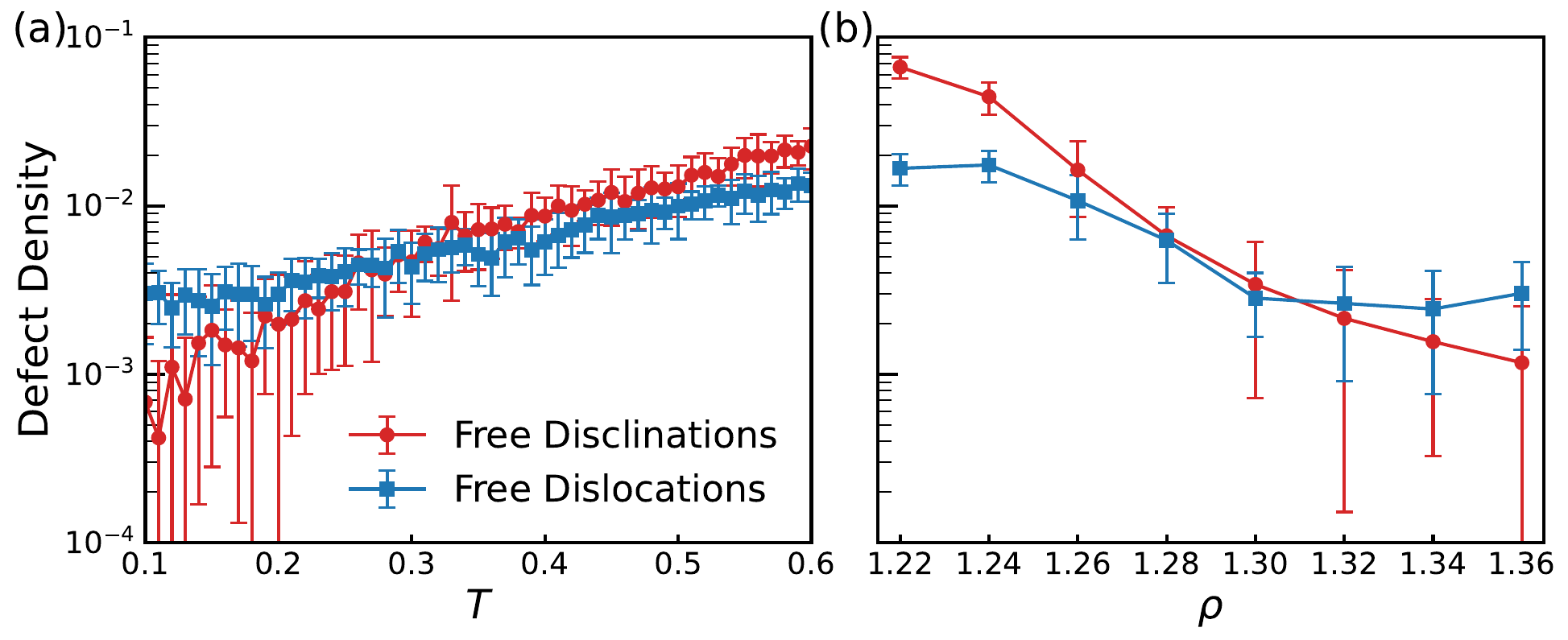}  
\caption{
Dependence of the defect density (a) on the temperature, at $\rho = 1.36$, and (b) on the density, at $T=0.1$.
Data are averaged over 10 independent simulations, and error bars represent the standard deviations.
\label{fig:defects}
}
\end{figure}
To confirm this picture microscopically, we analyze the density of topological defects in Fig.~\ref{fig:defects}(a–b), both at low temperature and under compression.
This analysis confirms that the system does not recover a solid-like order upon cooling or compression: dislocations remain at a finite density, whereas free disclinations remain suppressed. 
This defect pattern is characteristic of a hexatic, which thus remains stable as $T \to 0$ or the density increases.

\begin{figure}[!t]
  \centering
  \includegraphics[width=0.48\textwidth]{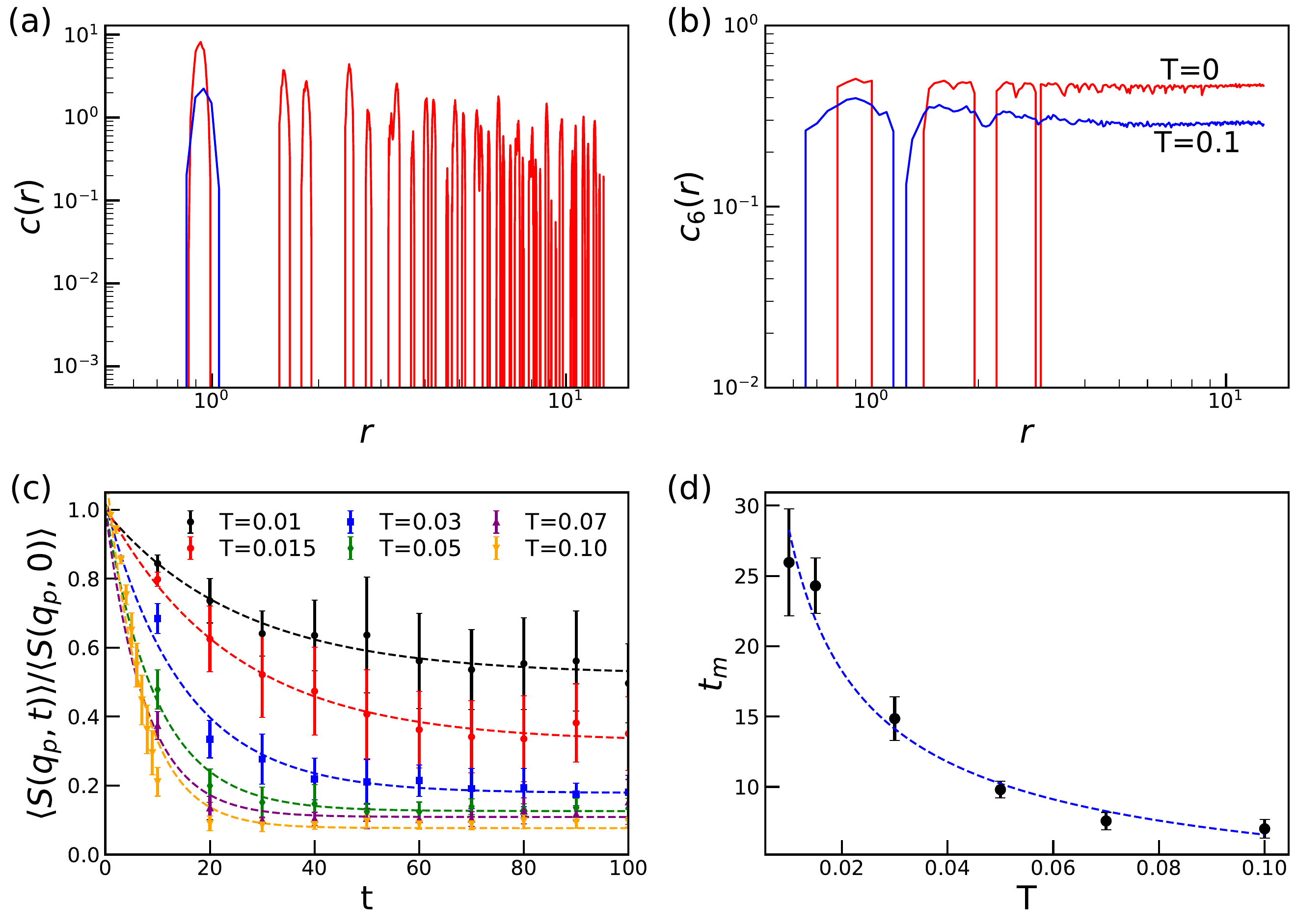}
  \caption{
  Translational (a) and bond-orientational (b) correlation function of an ideal solid at T = 0 (red), and of the configuration attained after heating it to $T=0.1$ (blue). Heating destroys positional order while retaining bond-orientational order.
  (c) Upon heating, the peak value of the static structure factor exponentially decays to an asymptotic low value on a melting time $t_m$.
  Error bars show standard deviations from ten independent runs.
  (d) Temperature dependence of $t_m$. The dashed line corresponds to $t_m \propto T^{-1/2}$.
  The uncertainties reflect the standard errors from the exponential fits.
  \label{crystal_melting}}
\end{figure}

To assess the robustness of the phase diagram—particularly the persistence of the hexatic phase at low temperatures and high densities—we verify that this phase also emerges upon heating a crystalline configuration prepared at zero temperature. 
Figs.~\ref{crystal_melting}a--b shows that, when the temperature is brought to  $T = 0.1$, the system loses translational order while retaining bond-orientational order, demonstrating that the hexatic phase is stable at low temperatures.
The same scenario occurs in larger systems, as we illustrate in the End Matter, implying that the stability of the hexatic phase is not due to the finite size of our sample.
To systematically characterize the temperature dependence of this melting process, we monitor the time evolution of the static structure factor $S(\vec{q}_{\rm p}, t)$ at one of the first Bragg peaks $\vec{q}_{\rm p}$. 
If the $T = 0$ crystal were metastable, melting would proceed via an activated mechanism, with $S(\vec{q}_{\rm p}, t)$ remaining constant up to a melting time $t_m$ scaling as $t_m \propto e^{E_m/T}$. 
In contrast, for an instability-driven melting process, $S(\vec{q}_{\rm p}, t)$ is expected to decay continuously on the vibrational time scale, which in the underdamped regime scales as $t_m \propto T^{-1/2}$. 
We validate this scenario in Fig.~\ref{crystal_melting}(c) and (d), showing that $S(\vec{q}_{\rm p}, t)$ approaches exponentially its asymptotic value, and that the characteristic melting time has the temperature dependence expected for fluctuation-driven melting. 
These results indicate that the crystalline phase is intrinsically unstable at any finite temperature. 
This finding is consistent with the equilibrium phase diagram shown in Fig.~\ref{phase_diagram}, and aligns with the expectation from Mermin--Wagner theory: in two dimensions, melting does not require activation over an energy barrier, but instead proceeds via continuous destabilization driven by long-wavelength thermal fluctuations.

Since the hexatic phase remains stable down to $T \to 0$, its free energy must be lower than that of the solid, $F_{\mathrm{hex}} < F_{\mathrm{sol}}$. 
For hard disks and regular polygons, this inequality arises purely from entropy: 
all configurations have zero energy, so stability follows from $S_{\mathrm{hex}} > S_{\mathrm{sol}}$ within the hexatic density range. 
In our system, $F_{\mathrm{hex}} < F_{\mathrm{sol}}$ at high densities, where the elastic energy of the compressed systems is finite,
demonstrating that the hexatic phase not only retains an entropic advantage but also achieves a lower or comparable energy, $E_{\mathrm{hex}} \leq E_{\mathrm{sol}}$.

\begin{figure}[!t]
\centering
\includegraphics[width=0.48\textwidth]{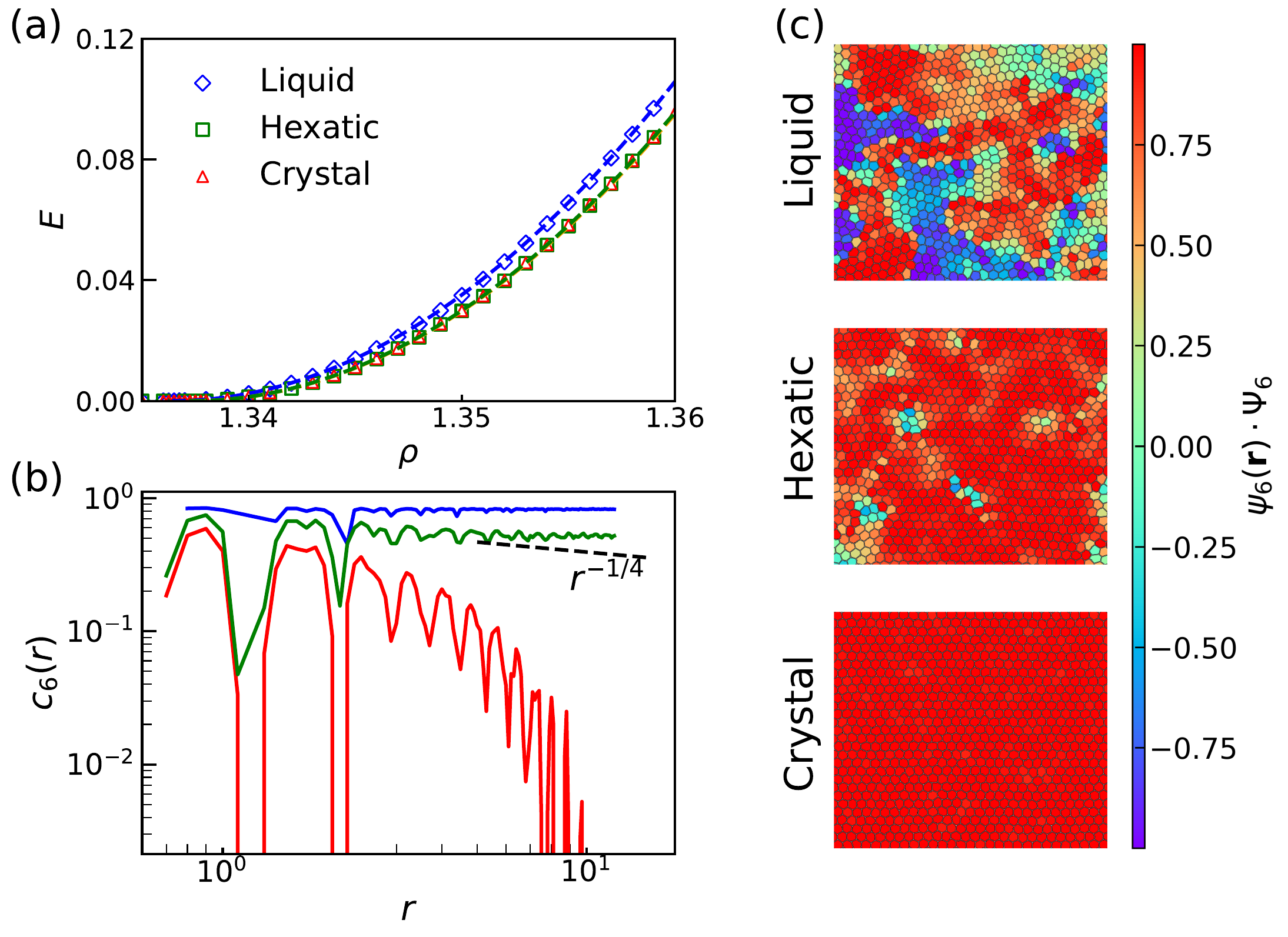}
\caption{                
(a) Density dependence of the elastic energy upon quasistatic compression, for systems prepared in liquid-like, hexatic-like, and solid phases. 
(b) The symmetries of the initial configuration are preserved during compression, as demonstrated by the bond-orientational correlation functions at the highest attained density.
(c) Representative snapshots of the system with particles colored according to the scalar product between the local sixfold bond-orientational order parameter and its average across all particles.        
\label{fig:jamming}
}
\end{figure}

To verify this prediction, we investigate the jamming behavior of the model.
In doing so, we extend the previous investigation by Boromand et al.~\cite{boromand2018jamming} to larger systems and, more importantly, to initial configurations with different spatial symmetries - liquid, hexatic, and crystalline.
Specifically, we prepare $T = 0$ configurations with liquid and hexatic symmetries by minimizing representative finite-temperature state points, and generate solid configurations by placing particles on a triangular lattice at low density.
We then quasistatically compress these configurations by incrementing the density by $\delta \rho = 2.5\cdot10^{-4}$ and minimizing the energy after each step. 
Fig.~\ref{fig:jamming}(a) illustrates the dependence of the energy of the system on the density.
Liquid-like configurations jam before the others and have a higher energy under compression. 
In turn, the energy of hexatic-like and liquid-like configurations is the same at all considered densities.
Importantly, these two phases behave identically despite retaining their original symmetries under compression.
This is demonstrated in Fig.~\ref{fig:jamming}(b), which demonstrates that the decay of the bond-orientational correlation function at the highest density we have achieved still exhibits the features associated with the different phases, and in Fig.~\ref{fig:jamming}(c), where we show snapshots of the systems at the highest considered density, with particles color coded according to the alignment between local and global bond-orientational order parameter.
These results clarify that at high density, in the $T \to 0$ limit $E_{\mathrm{hex}} \simeq E_{\mathrm{sol}} < E_{\mathrm{liq}}$.
This explains why the solid phase does not appear at finite temperature: without an energetic advantage, the system is stabilized in the hexatic phase by its higher entropy.

\begin{figure}[!!t]
    \centering
    \includegraphics[width=0.41\textwidth]{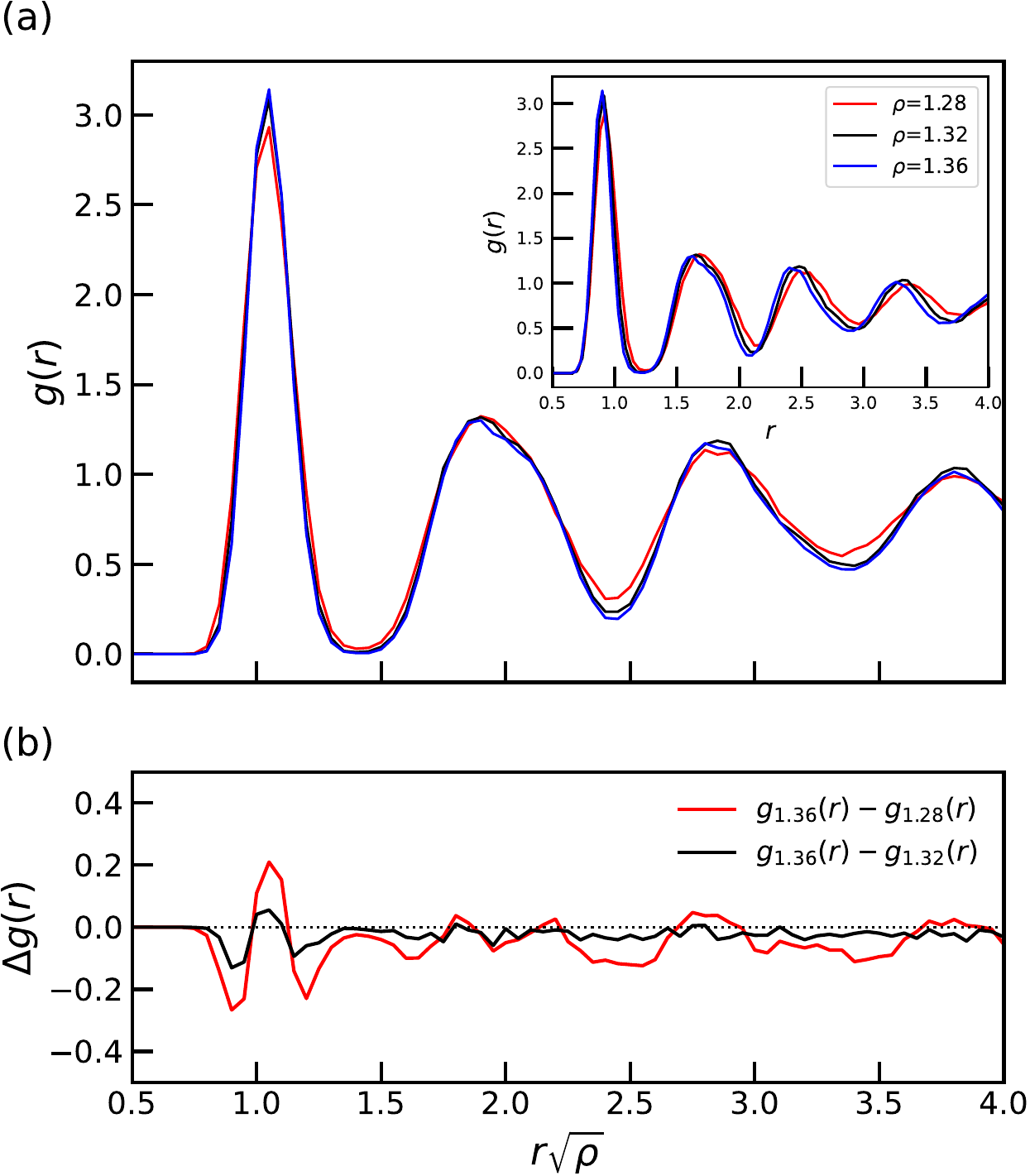}
    \caption{
    (a) Radial distribution functions in the hexatic phase at $T = 0.1$, plotted versus $r$ (inset) and versus the scaled distance $r\sqrt{\rho}$.
    Rescaling the distances induces a good collapse of the distribution at the two highest densities, as clarified in (b) via a plot of $\Delta g(r)$ versus the rescaled distance. Data are averaged over 10 independently generated configurations.
    }
    \label{fig:structure}
\end{figure}

The absence of a transition to the solid phase upon compression, both at zero and finite temperature, indicates that the system accommodates density increments without incurring the energetic cost that normally drives dislocation condensation and crystallization. To understand this response, we examine the density dependence of structural correlations. As shown in Fig.~\ref{fig:structure}(a,b), rescaling interparticle distances by $\sqrt{\rho}$ collapses the radial distribution functions of dense hexatic states onto a single curve, revealing predominantly affine particle displacements, as observed in soft microgels~\cite{Romeo2013}. This suggests that particle deformability localizes strain within individual particles and screens elastic interactions between dislocations.

To further substantiate the idea that particles accommodate compression through shape changes and strain localization, we decompose the elastic energy of jammed configurations (Fig.~\ref{fig:jamming}(a)) into area, perimeter, and interaction contributions (End Matter Fig.~\ref{fig:energy_decomposition}).
Compression primarily increases area and interaction energies, while the perimeter term remains negligible—typically two orders of magnitude smaller—indicating that particles accommodate strain by changing shape rather than by promoting the stress localization required for dislocation binding.
This mechanism is not generic to all soft-particle systems~\cite{Ciamarra2013c}: it relies on particles shrinking under compression, which prevents the formation of additional contacts and preserves the affine response at high density. 
Interestingly, while the area decreases with density, the perimeter distribution broadens slightly without significant changes in the mean (see Fig.~\ref{fig:dperi_histograms}), indicating that compression is accommodated by particle shrinkage and shape distortions, which suppresses the energetic drive for dislocation condensation and crystallization.

Our results demonstrate that deformable particles can stabilize a hexatic phase down to zero temperature and high density, effectively suppressing the conventional solid phase, in the absence of attractive interactions~\cite{pasupalak2024epithelial}. 
Deformability suppresses the energetic difference between the hexatic and solid phases, favoring the former at finite temperature due to its higher entropy. 
The effect relies critically on particle softness: we find that stiffer particles recover a more conventional melting scenario, as previously reported~\cite{Guo2023}. 
More broadly, our findings illustrate that particle deformability extends entropy-dominated behavior to finite-energy configurations, suggesting that similar mechanisms may govern phase behavior in soft and biological 2D systems, from microgels to epithelial tissues.

\begin{acknowledgements}
This research was supported by the Singapore Ministry of Education through grants
MOE-T2EP50221-0016 and RG152/23.
\end{acknowledgements}

\bibliography{citation}

\clearpage
\section*{End Matter}

\begin{figure}[!h]
    \centering
    \includegraphics[width=0.45\textwidth]{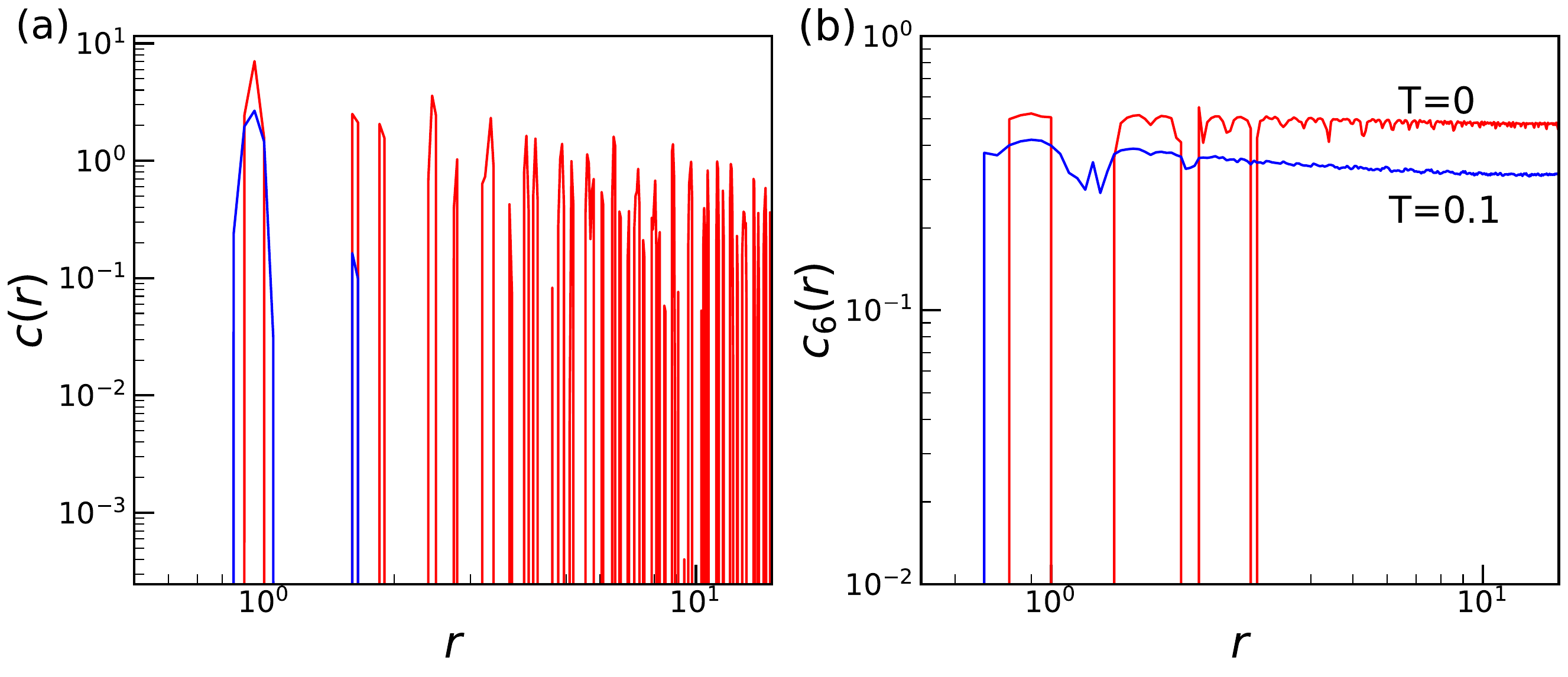}
    \caption{
    Translational (a) and bond-orientational (b) correlation function of an ideal crystal with 4096 cells at $T = 0$ (red), and of the hexatic configuration in which the crystal transitions to when the temperature is brought to $T=0.1$ (blue).    
     \label{fig:larger_sys}
    }   
\end{figure}

{\it Size effects -- } To demonstrate the robustness of our results to size effects, we focus on the temperature-induced melting of a crystalline configuration prepared at zero temperature. Fig.~\ref{crystal_melting}(a) and (b) illustrate that a system of $N=1024$ particles readily transitions from the crystal to the hexatic phase as temperature is brought to $T=0.1$. 
Fig.~\ref{fig:larger_sys} shows that the same results occur in a system with $N=4096$.

\begin{figure}[!h]
    \centering
    \includegraphics[width=0.45\textwidth]{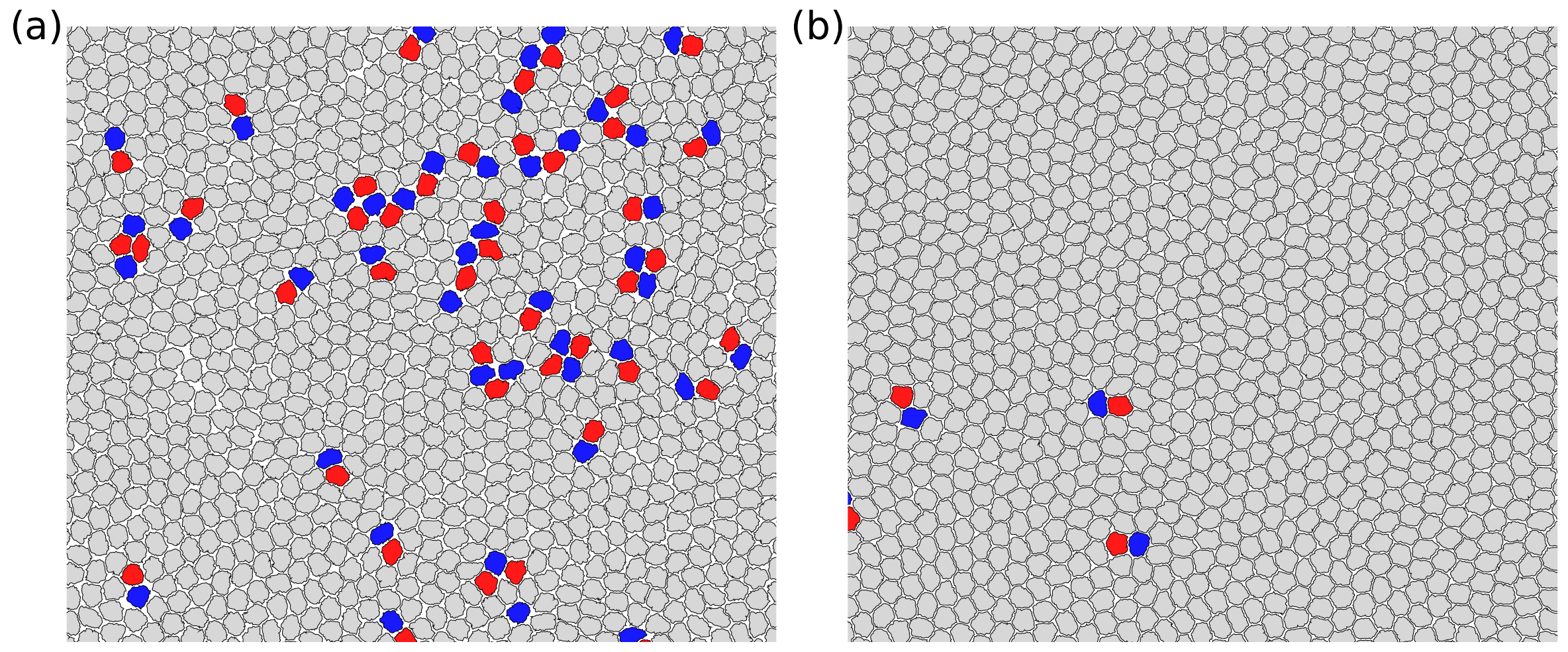}
    \caption{
    Representation of the system at $\rho = 1.32$ at temperatures $T = 0.65$ (a, liquid) and $T = 0.1$ (b, hexatic). 
    Particles with $n_c = 5,6$ and $7$ neighbors are illustrated in red, gray and blue, respectively.
    }
    \label{fig:defects2}
\end{figure}

{\it Defects -- } 
Fig.~\ref{fig:defects2} provides snapshots of the system in the hexatic and liquid phases, with particle colors indicating the number of nearest neighbors.
In the manuscript, we investigated the density of free disclinations, particles with $n_v = 5$ or $n_v = 7$ Voronoi neighbors, all of which have $n_v = 6$. 
Similarly, we investigated the density of free dislocations, defined as the density of adjacent $5$--$7$ defect pairs surrounded by particles with $n_v = 6$.

\begin{figure}[!h]
    \centering
    \includegraphics[width=0.48\textwidth]{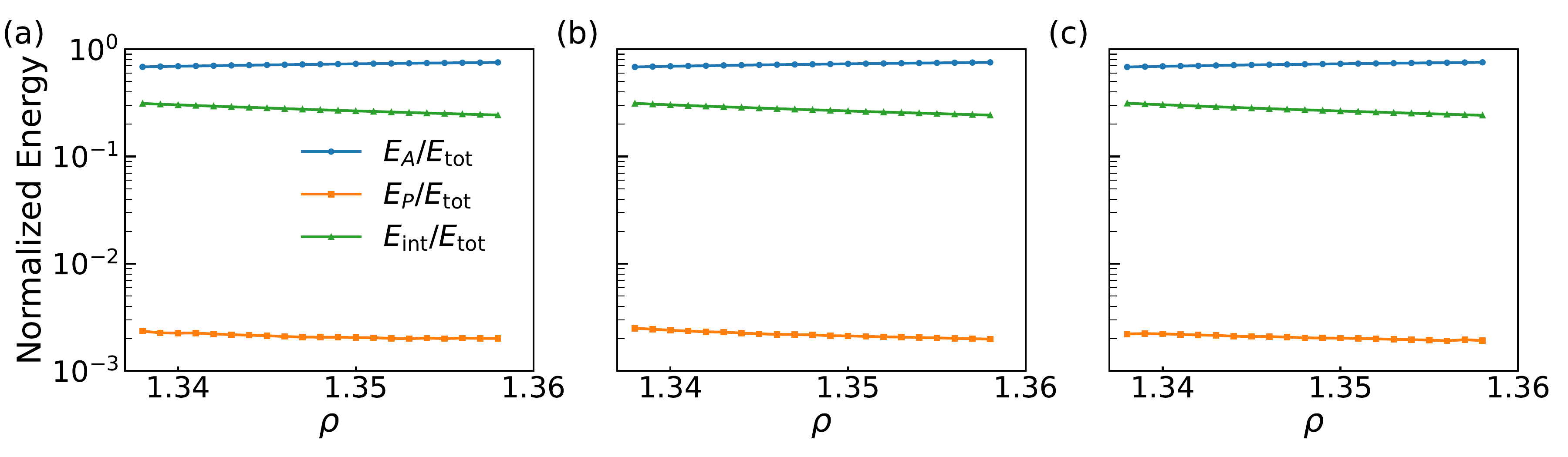}
    \caption{
    Density dependence of the fraction of the elastic energy $E$ of jammed configuration ($E > 0$) accounted for by the area, the perimeter, and the interaction term, for systems with liquid-like (a), hexatic-like (b), and solid-like symmetries (c). 
    }
    \label{fig:energy_decomposition}
\end{figure}

\begin{figure}[!h]
    \centering
    \includegraphics[width=0.48\textwidth]{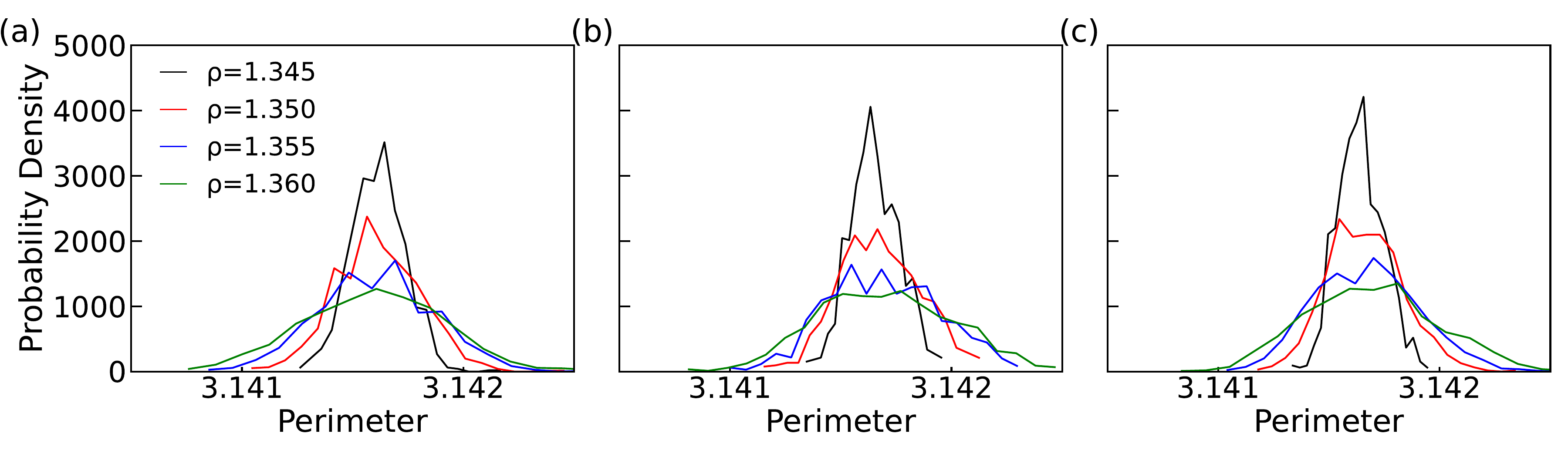}
    \caption{
    Probability distribution of the perimeter at selected densities, as the system is compressed at zero temperature above the jamming onset,  for systems with liquid-like (a), hexatic-like (b), and solid-like symmetries (c). 
    }
    \label{fig:dperi_histograms}
\end{figure}

{\it Aria, perimeter and interaction energies -- } 
We decompose the elastic energy of jammed configurations as
$E = E_A + E_P + E_{\rm int}$,
where $E_A = \sum_\alpha K_A (A_\alpha - A_0)^2$ is the area contribution, with $\alpha$ running over all particles;
$E_P = \sum_\alpha \sum_i k_p (l_i - l_0)^2$ is the perimeter contribution, where the inner sum runs over all bonds $i$ of particle $\alpha$;
and $E_{\rm int} = \sum_{ab} u(r_{ab})$ is the interaction contribution, where the sum runs over all pairs of monomers belonging to distinct particles.
Figure~\ref{fig:energy_decomposition} shows that in jammed configurations, the perimeter term contributes a negligible fraction of the total energy, indicating that shape changes accommodate strain at almost no energetic cost.
Consistent with this, Fig.~\ref{fig:dperi_histograms} demonstrates that the average perimeter distribution does not change sensibly with the density, while its distribution slightly broadens, implying that cells distort to keep a nearly constant perimeter as their area decreases under compression. 

\end{document}